\documentclass{radioeng}

\usepackage{graphicx} 

\usepackage{amsmath,amsfonts}%
\usepackage{ntheorem}%
\usepackage[title]{appendix}%
\usepackage{xcolor}%
\usepackage{textcomp}%
\usepackage{manyfoot}%
\usepackage{booktabs}%
\usepackage{algorithm}%
\usepackage{listings}  
\usepackage{multirow}%
\usepackage[backend=biber, isbn=true]{biblatex}
\theoremstyle{plain}
\newtheorem{theorem}{Theorem}
\renewtheorem{theorem*}{Proof}

\newcommand{\figdir}{./images}

\newcommand{\beq}{\begin{equation}}
	\newcommand{\eeq}{ \end{equation} }
\newcommand{\bea}{\begin{eqnarray}}
	\newcommand{\eea}{\end{eqnarray}}
\newcommand{\ba}{\begin{array}}
	\newcommand{\ea}{\end{array}}
\newcommand{\no}{\nonumber}

\newcommand{\diag}{\ensuremath{\text{diag}}}

\newcommand{\fig}[1]{Fig.~\ref{#1}}

\newcommand{\float}[3]{\ensuremath{{#1},\kern-0.12em{#2}\cdot 10^{#3}}}
\newcommand{\double}[2]{\ensuremath{{#1},\kern-0.12em{#2}}}

\newcommand{\abs}[1]{\ensuremath{\left\lvert{#1} \right\rvert}}
\newcommand{\R}{\ensuremath{\mathbb{R}}}
\newcommand{\C}{\ensuremath{\mathbb{C}}}

\newcommand{\E}{{\rm I\kern-.3em E}}

\newcommand{\area}{
CIRCUITS
} 

\begin{document}


\setcounter{firstpage}{1}


\rengTitle{The Impact of the Cauchy Interlace Theorem on the 
	Convergence of the RLS Algorithm}

\rengNames{Hans~Georg BRACHTENDORF $^\mathit{1}$}

\rengAffil{$^1$ Signal Processing Laboratory, University of Applied Sciences of Upper Austria, 
	Softwarepark 11, 4232 Hagenberg, Austria}

\rengMail{hans-georg.brachtendorf@fh-hagenberg.at}

\rengReceived{September 1, 2026}{September 1, 2026} 

\begin{multicols}{2}

\begin{rengAbstract}
The recursive least squares (RLS) algorithm is widely used for the adaptive
equalization of fading channels. Basically, the technique behind the RLS algorithm 
is a fast update of the
inverse autocorrelation matrix with ${\mathcal{O}}(N^2)$ computational
complexity. The (inverse) autocorrelation
matrix is normally initialized proportionally to the identity matrix,
leading to clustered eigenvalues.

In this paper, we
revisit the convergence of the RLS algorithm in view of the Cauchy Interlace Theorem.
It states that the eigenvalues of a Hermitian matrix $A$
and of the rank one updated matrix $\hat{A} = A + u\,u^H$
interlace. We analyze the impact
of this theorem on the convergence of the RLS algorithm and its suitable 
initialization. The results suggest an improved initialization for speeding up
the convergence time.
Moreover, a novel proof
of the theorem is given based on the Sherman Morrison Woodbury formula.
\end{rengAbstract}

\rengKeywords{RLS algorithm, RLS convergence,
	Cauchy Interlace Theorem (CIT), settling time, 
	Sherman Morrison Woodbury formula}


\rengSection{Introduction}
\label{sec1}

The RLS algorithm for channel estimation
belongs to the classical algorithms in signal processing. 
It is known that it converges must faster than the
least mean squares (LMS) algorithm on the expense of a significantly
higher numerical cost (${\mathcal{O}}(N^2)$ instead of ${\mathcal{O}}(N)$).
For brevity, we
reference only a view well-known textbooks 
\cite{haykin2002adaptive,Sayed2008,kammeyer2017nachrichtenUebertragung,widrow1985adaptive,Mert01},
offering a thorough derivation and discussion. It is considered
that the convergence speed depends solely on the forgetting factor,
a free parameter chosen less than or equal unity. It is known, that
a tradeoff exists between convergence speed, tracking of time-variant
channels and sensitivity to noise, depending on the choice of 
the forgetting factor.

For the initialization of the RLS algorithm, 
one needs further an estimate of the inverse of the autocorrelation matrix of the received signal. 
In this paper we analyze the impact of this initialization w.r.t.\
the convergence or settling time of the RLS algorithm. We show that this problem is tightly linked to the Cauchy Interlace Theorem (CIT) from linear algebra. In contrast to the standard assumptions that
the convergence depends solely on the forgetting factor, it can be shown
that the initialization is cruical and standard assumptions are only
partly true.

The Cauchy Interlace Theorem \cite{golub1996matrix}
states that
the eigenvalues of a Hermitian matrix $A$ and of a rank one updated
matrix $\hat{A} = A + u\,u^H$ with column vector $u$ interlace.
We show that this theorem has a tremendous impact on the settling time for the
adaptive RLS algorithm. Notably, the standard initialization, initializing
the (inverse) autocorrelation
matrix proportional to the identity matrix, is the worst of all choices, since the eigenvalues
are clustered. Clustered eigenvalues, as shown in this paper, slow down
the convergence significantly, which is a consequence of Cauchy's theorem.
Furthermore, a novel proof of the CIT theorem based on the
Sherman Morrison Woodbury formula is given in the appendix.

This paper is organized as follows: in Section~\ref{sec2} the RLS algorithm is revisited briefly, mainly for introducing the notation used throughout this paper. 
Besides the classical formulation of RLS, which exhibits an implicit scaling (bias), a scaling-free
formulation is introduced, which is the basis of our analysis. 
In Section~\ref{sec3} we investigate 
the settling/convergence time of this algorithm.
State of the art analyses consider solely the asymptotic behavior,
when the iteration count is much larger than the order of the transversal
filter, neglecting the impact of the initialization. These results
are therefore only partly true.

Here,
in view of the Cauchy Interlace Theorem, the convergence behavior is
studied when the iteration count is small and the initialization
comes into play. The findings
are verified by four typical channels.
A best practice rule for initializing the RLS algorithm is suggested.
The appendix gives also a novel proof of the CIT,
which is cruical for considering the convergence behavior, based on
the Sherman Morrison Woodbury formula.

\newpage

\rengSection{The RLS algorithm}
\label{sec2}

This section revisits briefly the RLS algorithm for notational purposes mainly. We refer also
to the vast literature on this topic 
\cite{haykin2002adaptive,Sayed2008,kammeyer2017nachrichtenUebertragung,widrow1985adaptive,Mert01}.
The standard formulation exhibits a bias of the estimated correlation,
here characterized by  the \emph{tilde} symbol, e.g., $\tilde{R}_{xx}$. It
cancels out in practical applications of the RLS algorithm and is normally not
mentioned in the vast literature on this topic. However, for the convergence analysis
presented in this paper, a scaled version is of advantage, characterized by the \emph{hat} symbol,
e.g., $\hat{R}_{xx}$.

Given a discrete time system with a
data source $a[k] \in \mathcal{A}$ from a constellation diagram
$\mathcal{A}$, a channel impulse response $h$, additive Gaussian noise $n$, received
signal $x$, an equalizer with impulse response $g$ and equalizer output signal $y$.
The data sequence $a$ and noise sequence $n$ are usually
assumed to be statistically independent sequences with zero mean and variances $\sigma_a^2$ and
$\sigma_n^2$, respectively,
and vanishing cross-correlation. 
Employing the notation
\begin{alignat*}{1}
	\mathbf{h} &= ( {h}[0],\ldots,\,{h}[K])^T \in \C^{K+1} \\
	\mathbf{g} &= ( {g}[0],\ldots,\,{g}[N])^T \in \C^{N+1}\\
	\mathbf{x}[k] &= ( {x}[k],\ldots,\,{x}[k-N])^T \in \C^{N+1} \\
	\mathbf{n}[k] &= ( {n}[k],\ldots,\,{n}[k-N])^T \in \C^{N+1}\\
	\mathbf{a}[k] &= ( {a}[k],\ldots,\,{a}[k-N-K])^T \in \C^{K+N+1} \\
\end{alignat*}
where $K$ is the order of the channel $h$, $N$ the order of the 
equalizer impulse response $g$, $a$ the stimulus, $n$ the additive noise
(assumed zero-mean and uncorrelated) and $x$ the received signal.
Moreover
\bea
H = 
\left(
	\begin{array}{cccccccc}
		h[0]    &          &          & && &&\\
		h[1]    & h[0]     &          & && &&\\
		h[2]    & h[1]     & h[0]     & && &&\\
		\vdots  & \vdots   & \vdots   & && &&\\
		h[K]    & h[K-1]   & h[K-2]   & && &\ddots&\\
		& h[K]     & h[K-1]   & && &\ddots&h[0]\\
		&          & h[K]     & && &\ddots&h[1]\\
		&          &          & && &&h[2]\\
		&          &          & && &\ddots&  \vdots\\
		&          &          & && &&  h[K]
	\end{array}
	\right)  \no
	\eea
	where $H \in  \C^{N+K+1 \times N+1}$ is the impulse response matrix of the channel. 
	The autocorrelation matrix $R_{xx} \in \C^{N+1 \times N+1}$
	is given by
	\[
	R_{xx} = \E\left\{\mathbf{x}^*\,\mathbf{x}^T \right\}
	= \sigma_a^2 \; \left(H^H\,H + \frac{\sigma_n^2}{\sigma_a^2}\, I \right)
	\]
	where $^H$ is the Hermitian operator, the asterisk stands for
	conjugation and $\E$ means expectation.
	The vector of cross correlations is given by
	\[
	\mathbf{r}_{xa} = \E\{\mathbf{x}^*\,a[k-k_0]\} = \sigma_a^2\, H^H\,e_{k_0}
	\]
	with the canonical vector
	\[
	\no e_{k_0} = \left(0,\,\ldots,\,0,\,\underbrace{1}_{k_0+1},\,0,\,\ldots,\,0\right)^T
	\]
	and free variable $k_0 \in [0,\,N]$, the delay parameter.
	The optimal minimum mean square error (MMSE) equalizer coefficients are calculated by
	\[
	\mathbf{g}_{\mathrm{opt}} = R_{xx}^{-1}\,\mathbf{r}_{xa} =
	\left(H^H\,H + 
	\frac{\sigma_n^2}{\sigma_a^2}\, I \right)^{-1}\,H^H\,e_{k_0}
	\]
	The expectation values are estimated in practice from the received signal by
	the unbiased estimator
	\begin{align*}
		\hat{R}_{xx}[k] &= c[k] \sum_{i=0}^k \mu^{i}\,\mathbf{x}^*[k-i]\,\mathbf{x}^T[k-i]\\
		\hat{\mathbf{r}}_{xa}[k] &= c[k] \sum_{i=0}^k \mu^{i}\,\mathbf{x}^*[k-i]\,a[k-k_0-i]
	\end{align*}
	with an adjustable parameter $0<\mu<1$, the forgetting factor, and scaling factor
	\[
	c^{-1}[k] = \sum_{i=0}^k \mu^{i} = \frac{1-\mu^{k+1}}{1-\mu}
	\]
	which can be recursively calculated by $c^{-1}[k+1] = 1 + \mu\,c^{-1}[k]$.
	In practical implementations of the RLS algorithm the scaling factor $c[k]$ can be omitted
	because it cancels out in the calculation of the adaptive filter coefficients
	\beq
	\hat{\mathbf{g}}[k] = \hat{R}^{-1}_{xx}[k]\,\hat{\mathbf{r}}_{xa}[k]
	\nonumber
	\eeq
	
	By introducing the short hand $\nu_{k} = \frac{\mu\,c^{-1}[k]}{1+\mu\,c^{-1}[k]}$
	one obtains the recursions
	\begin{align*}
		\hat{R}_{xx}[k+1] &= \nu_{k} \, \hat{R}_{xx}[k] + 
		(1-\nu_{k})\,\mathbf{x}^*[k+1]\,\mathbf{x}^T[k+1]\\
		\hat{\mathbf{r}}_{xa}[k+1] &= \nu_{k} \,\hat{\mathbf{r}}_{xa}[k] +
		(1-\nu_{k})\,\mathbf{x}^*[k+1]\,a[k-k_0+1]
	\end{align*}
	with $\nu[k] \stackrel{k\rightarrow \infty}{\rightarrow} 1$.
	Without scaling factor the recursions read
	\begin{align*}
		\tilde{R}_{xx}[k+1] &= \mu \, \hat{R}_{xx}[k] + 
		\mathbf{x}^*[k+1]\,\mathbf{x}^T[k+1]\\
		\tilde{\mathbf{r}}_{xa}[k+1] &= \mu \,\hat{\mathbf{r}}_{xa}[k] +
		\mathbf{x}^*[k+1]\,a[k-k_0+1]
	\end{align*}
	which is the common practice in implementation.
	The RLS algorithm employs the Sherman Morrison Woodbury (SMW) formula\footnote{The SMW formula
		is often referred to as the matrix inversion lemma.} for recursively
	calculating the estimate $\mathbf{\hat{g}}$ with ${\mathcal{O}}(N^2)$ complexity.
	The Sherman Morrison Woodbury formula states that
	\[
	(A + u\, u^H)^{-1} = A^{-1} - A^{-1}\,u \, c \, u^H\,A^{-1},\quad
	c = \frac{1}{1 + u^H\,A^{-1}\,u}
	\]
	assuming that $A$ and $A + u\cdot u^H$ are regular matrices\footnote{The formula
		will be used below in the appendix for proving the interlace theorem too.}.
	By defining the Kalman gain
	\begin{align*}
		\tilde{\ell}[k+1] &:= \tilde{R}^{-1}_{xx}[k+1]\,\mathbf{x}^*[k+1] \\
		&=
		\frac{\mu^{-1}\,\tilde{R}^{-1}_{xx}[k]\,\mathbf{x}^*[k+1]}{1 + \mu^{-1}
			\,\mathbf{x}^T[k+1]\,\tilde{R}^{-1}_{xx}[k]\,\mathbf{x}^*[k+1]}
	\end{align*}
	one obtains the recursion for the correlation matrix
	\[
	\tilde{R}^{-1}_{xx}[k+1] = \mu^{-1}\,\tilde{R}^{-1}_{xx}[k]
	- \mu^{-1}\,\tilde{\ell}[k+1]\,\mathbf{x}^T[k+1]\, \tilde{R}^{-1}_{xx}[k]
	\]
	The estimated equalizer coefficients are recursively updated by
	\[
	\tilde{\mathbf{g}}[k+1] = \tilde{\mathbf{g}}[k] + \tilde{\ell}[k+1]\,\left( a[k-k_0+1] 
	-\mathbf{x}^T[k+1]\, \tilde{\mathbf{g}}[k] \right)
	\]
	Alternatively, taking the scaling $\nu_k$ into account
	\begin{align*}
		{\ell}[k+1] &= \hat{R}^{-1}_{xx}[k+1]\,\mathbf{x}^*[k+1] \\
		&=
		\frac{\nu_k^{-1}\,\hat{R}^{-1}_{xx}[k]\,\mathbf{x}^*[k+1]}{1 + (1-\nu_k)\,\nu_k^{-1}
			\,\mathbf{x}^T[k+1]\,\tilde{R}^{-1}_{xx}[k]\,\mathbf{x}^*[k+1]}
	\end{align*}
	yields
	\[
	\hat{R}^{-1}_{xx}[k+1] = \nu_k^{-1}\,\hat{R}^{-1}_{xx}[k]
	- (1-\nu_k)\,\nu_k^{-1}\,{\ell}[k+1]\,\mathbf{x}^T[k+1]\, \hat{R}^{-1}_{xx}[k]
	\]
	and
	\begin{align*}
		&\hat{\mathbf{g}}[k+1]  = \hat{\mathbf{g}}[k] + 
		{\ell}[k+1]\,\left( a[k-k_0+1] 
		-\mathbf{x}^T[k+1]\, \hat{\mathbf{g}}[k] \right)
	\end{align*}
	Besides the forgetting factor $\mu$ and the delay parameter $k_0$, the
	initial autocorrelation matrix $\tilde{R}_{xx}[0]$ must be suitably chosen as
	considered next.

\rengSection{The impact of the Cauchy interlace theorem on the RLS algorithm}
\label{sec3}
\vspace{1em}

The standard initialization of the inverse autocorrelation matrix
is usually chosen proportionally to the identity matrix $I$, i.e.
\[
\tilde{R}_{xx}^{-1} = \delta^{-1}\,I
\]
with suitably chosen $\delta$, i.e., $N+1$ clustered eigenvalues.
The  usual initialization of the filter coefficients is given by\footnote{
	Note the double indexing of $\tilde{g}$: $n$ is the index
	of a vector element $\tilde{g}$ of length $N+1$. The second
	index $k$ refers to the iteration count, starting with $k=0$.}
\[
\mathbf{\tilde{g}}[n][0] =
\begin{cases}
	0 & \text{if}\quad n \neq k_0\\
	1 & \text{if}\quad n = k_0
\end{cases},
\quad n=1,\dots,N
\]
and $\mathbf{\tilde{r}}_{xd} = {\tilde{R}}^{-1}_{xx}\,\mathbf{\tilde{g}}$.

The convergence of the RLS algorithm has been studied extensively for
the asymptotic behavior, where the iteration count is much larger than
the number of filter taps, i.e., $k \gg N+1$. In 
\cite{haykin2002adaptive} Sec.~9.7 the mean-square error $\E\left\{\abs{\epsilon}^2\right\}$ for a forgetting factor $\mu = 1$
is asymptotically derived as
\[
  \E\left\{\abs{\epsilon[k]}^2\right\} = \frac{1}{k}\,\sigma_n^2\,
  \sum_{i=1}^{N+1}\frac{1}{\lambda_i},\quad k \gg N
\]
where $\sigma_n^2$ is the variance of a zero-mean additive noise
process and
the $\lambda_i$ are the eigenvalues of the autocorrelation matrix $R_{xx}$.
The derivation explicitly neglects the initialization term and is thus
only valid asymptotically for large $k$. The result shows that
the expectation of the squared error
vanishes for $k \rightarrow \infty$ with $1/k$ and that the convergence
depends asymptotically on the eigenvalue distribution
of the channel correlation
matrix.

In case of a time-varying
channel, modeled as a 1st.\ order Markov chain,  process noise
autocorrelation matrix $R_{\omega \omega}$ and forgetting factor $\mu < 1$ 
the  mean-square error
is asymptotically \cite{haykin2002adaptive}~Chap.~14
\[
 \E\left\{\abs{\epsilon[k]}^2\right\} \approx
 \frac{1-\mu}{2}\,\sigma_n^2\,\text{tr}\{R_{xx}^{-1}\} +
 \frac{1}{2(1-\mu)}\,\text{tr}\{R_{\omega \omega}\},\; k \rightarrow \infty
\]

when $k$ tends to infinity,
where $\text{tr}$ is the trace of a matrix. Note that
\[
 \text{tr}\{R_{xx}^{-1}\} = \sum_{i=1}^{N+1} \frac{1}{\lambda_i}
\]
where the $\lambda_i$ are again the eigenvalues of the 
correlation matrix.

For quasi-stationary channels
the second term vanishes, since $R_{\omega \omega} \rightarrow 0$.
Then the optimal choice for the forgetting factor
is $\mu = 1$, resulting asymptotically in a
vanishing mean square error.
These asymptotic results neglect the convergence behavior for small
iteration count $k$, which depends on the initialization of $R_{xx}$.

\begin{figure}
	\centering
	\includegraphics[width=0.99\columnwidth,keepaspectratio]{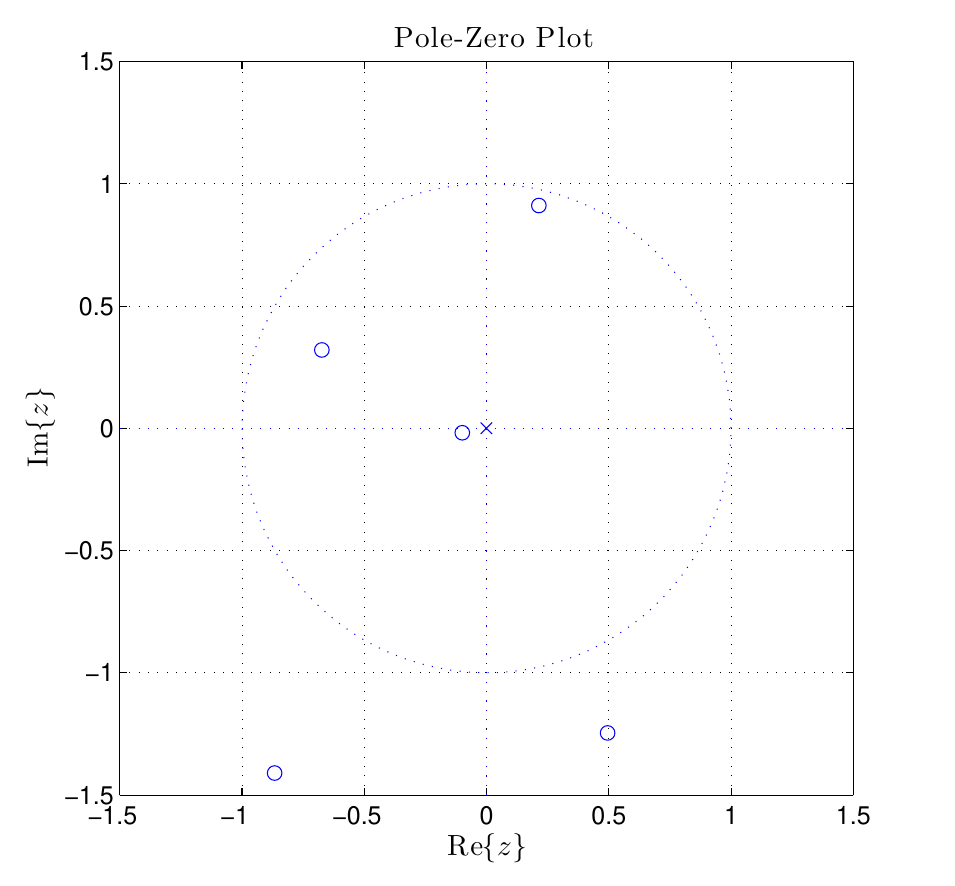}
	\fcaption{Pole zero plot of the mixed phase channel four.\label{\figdir/ChannelMixedPhase.pdf}}
\end{figure}
\vspace{1em}

For the simulations below the training symbols are chosen statistically independent
and identically distributed from the QPSK constellation diagram. 
Four different channels have been chosen:
$H_1(z) = 1 -z^{-1}$ with a zero of the transfer function $H_1(z)$ at $z_0=1$,
a maximal phase system $H_2(z) = (1 -0.9\,z^{-1})\cdot \exp{(j\,\pi/4)}$,
the allpass system\footnote{The infinite impulse response has been limited
	to $30$ tabs.}
$H_3(z) = \frac{1 - 1/z_{pol}\,z^{-1}}{1 - z_{pol}\,z^{-1}}$
and the mixed phase channel $H_4$ with poles and zeros depicted in \fig{\figdir/ChannelMixedPhase.pdf}.
Moreover, the channel response coefficients are normalized to $||h||_2 = 1$.
The initialization of $\tilde{R}_{xx}$ is done by
\[
\tilde{R}_{xx} = \diag\{10^a,\dots, 10^b\}
\]
where the initial eigenvalues are ordered in logarithmic scale. 
In what follows the results are differentiated by the tupel $(a,\,b)$,
i.e.\ $(a,\,b) = (\log_{10}(\delta),\,\log_{10}(\delta))$
corresponds to the standard
initialization of clustered eigenvalues $\lambda=\delta$.
The results of
$500$ statistically independent training sequences have
been averaged. The equalizer's transfer function is of order $N=32$ and the
forgetting factor $\mu=0.99$.
For depicting the settling curve a suitable filter tab
in the neighborhood of $k_0=14$, i.e.\ $g[k_0\pm 1]$, has been
chosen\footnote{The choice of $k_0$ depends on the channel characteristic,
i.e.\ maximal/minmal or mixed phase,
see the comprehensive literature on this subject.}. 
In the following figures the $90\%$ threshold of the asymptotic
value obtained from the MMSE solution is marked in black. 
Moreover, the fixed delay $k_0$ of the settling
time has been omitted in the plots.

\begin{figure}
	\centering
	\includegraphics[width=0.99\columnwidth,keepaspectratio]{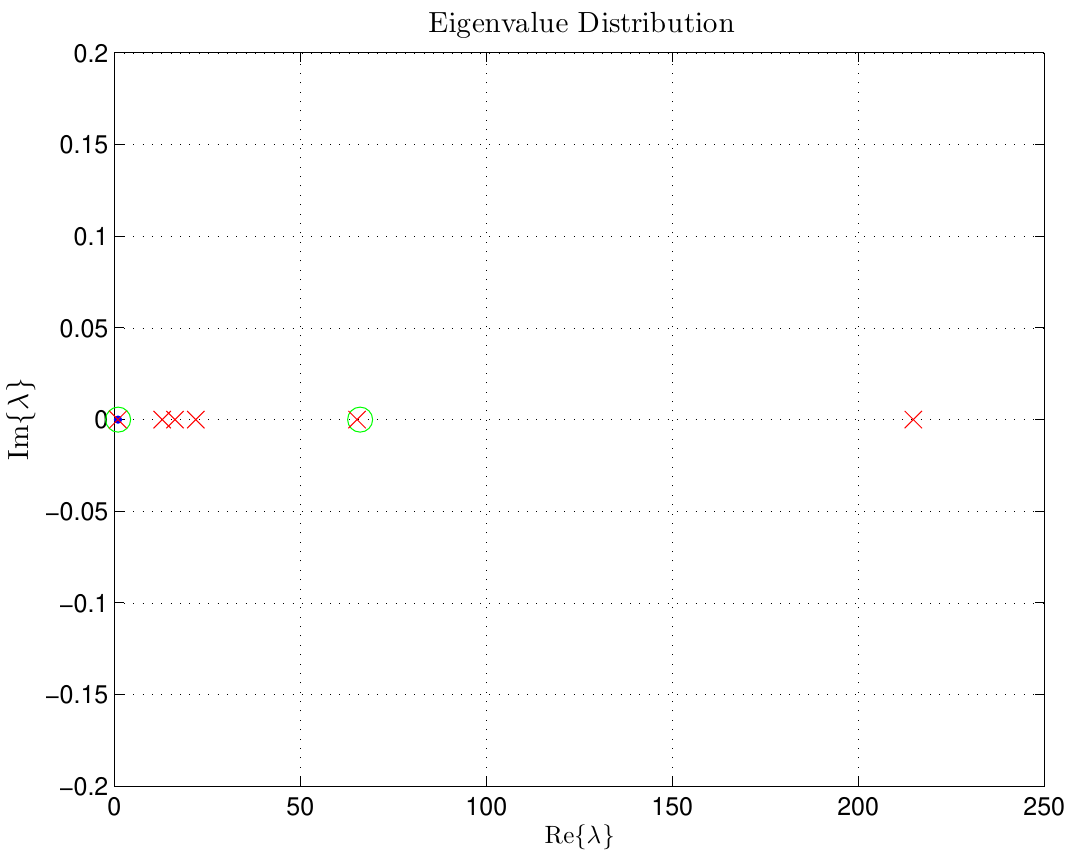}
	\fcaption{Eigenvalue distribution 
		for channel~$1$, initial ($\cdot$), 1~st iteration ($o$) 
		and 5~th iteration ($x$) with initialization $(a,\,b) = (0,\,0)$.\label{\figdir/EigenvalueDistribution.pdf}}
\end{figure}
\vspace{1em}

The \fig{\figdir/EigenvalueDistribution.pdf} shows the eigenvalue distribution of 
$\hat{R}_{xx}$ initialized with the identity matrix. They are plotted for the
for the initial $k=0$ ($\cdot$), $k=1$ ($o$) and $k=5$ ($x$) iteration.
One observes that the eigenvalues are clamped near $\lambda=1$ which is a direct
implication of the CIT\footnote{The cluster of
eigenvalues is slightly shifted by the forgetting factor,
i.e.\ $\lambda[k] = \mu^k$.}. At the $k$~th.\ iteration only $k$ eigenvalues can be updated,
the remaining ones stay fixed at $\lambda=\delta$. This must have obviously a tremendous
impact on the convergence speed.	
Moreover, this effect cannot directly be influenced by
the forgetting parameter $\mu$. Hence, the convergence towards the correct eigenvalues
of ${R}_{xx}$ is hence expected to be very slow.
Therefore it is expected that a thorough choice
of the initial autocorrelation matrix will have a strong impact on the 
convergence speed.
The plots below show the curves for the tupels $(a,\,b) = (0,\,0),\,(-2,\,0)$
and $(-3,\,0)$.

\begin{figure}
	\centering
	\includegraphics[width=0.99\columnwidth,keepaspectratio]{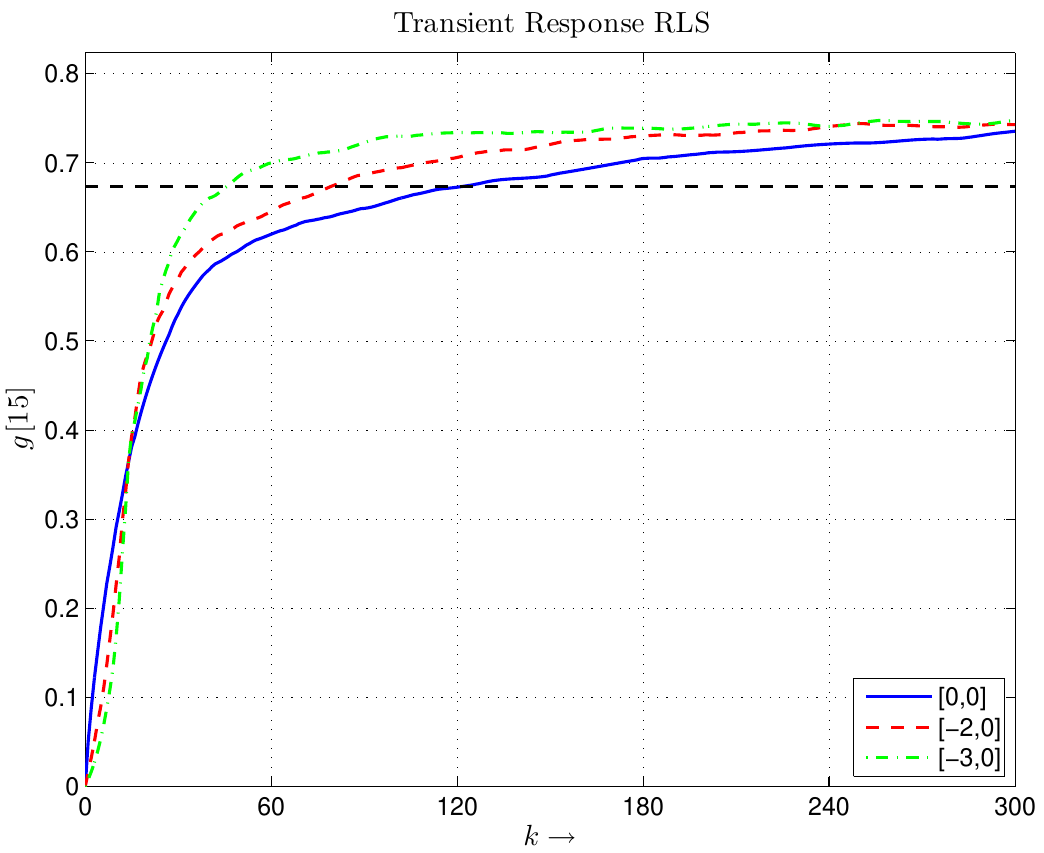}
	\fcaption{Initial transient response for channel~$1$ with initialization
		$(a,b) = \{(0,0),\,(-2,0),\,(-3,0)\}$.}\label{\figdir/Channel1.pdf}
\end{figure}
\vspace{1em}

\paragraph{Channel $H_1$}
For the channel~1 we obtain the extreme eigenvalues of the autocorrelation
matrix $\lambda_{max} = 1.99574,\, \lambda_{min} = 0.0043$, i.e.,
a ratio of $467.84$. \fig{\figdir/Channel1.pdf} depicts the averaged
transient response of filter coefficient $g[15]$. For the tupel
$(a,\,b) = (-3,\,0)$ one gets an improvement of the settling time
of more than a factor of $2$ compared with an initialization by the identity
matrix.

\begin{figure}
	\centering
	\includegraphics[width=0.99\columnwidth,keepaspectratio]{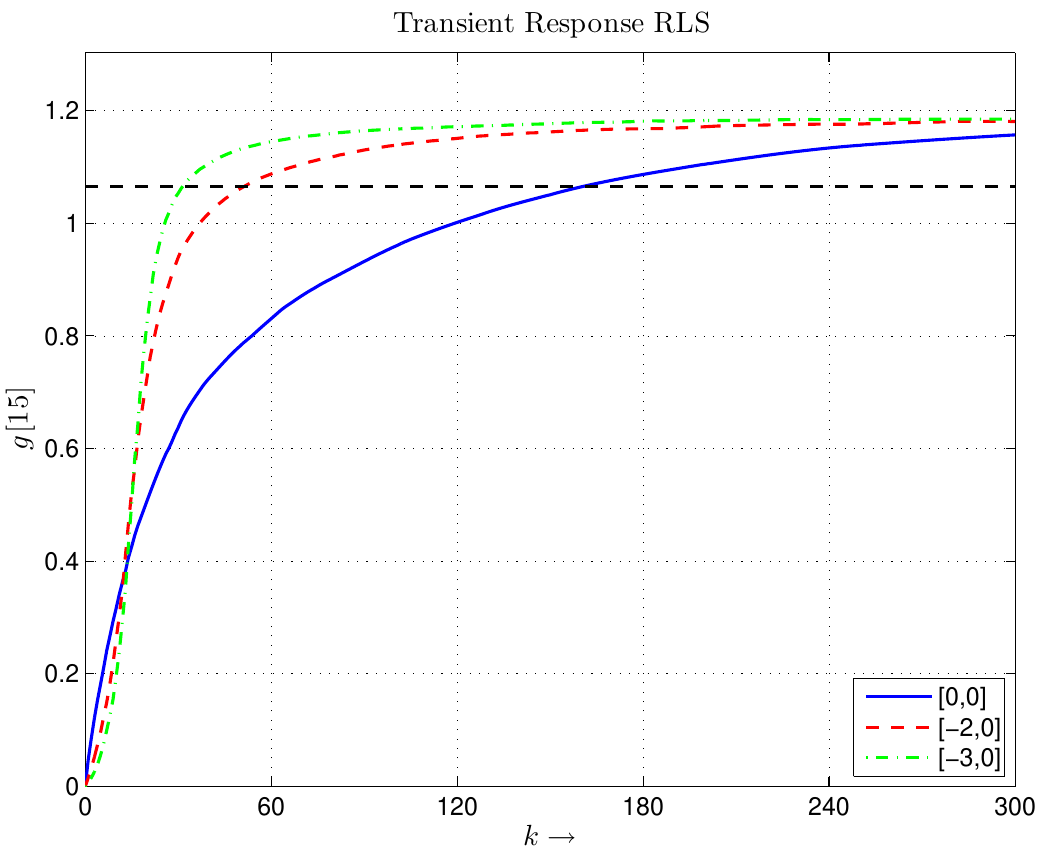}
	\fcaption{Initial transient response for channel~$2$ with initialization
		$(a,b) = \{(0,0),\,(-2,0),\,(-3,0)\}$.}\label{\figdir/Channel2.pdf}
\end{figure}
\vspace{1em}

\paragraph{Channel $H_2$}

For the maximal phase channel~2 the extremal eigenvalues read
$\lambda_{max} = 1.9902, \,  \lambda_{min} = 0.0098$, i.e.,
a ratio of $203.76$. The averaged transient responses are
depicted in \fig{\figdir/Channel2.pdf}. The settling time
for the parameter setting $(a,\,b) = (-3,\,0)$ is more than 
a factor of $4$ shorter than for an initialization by the identity matrix.

\begin{figure}
	\centering
	\includegraphics[width=0.99\columnwidth,keepaspectratio]{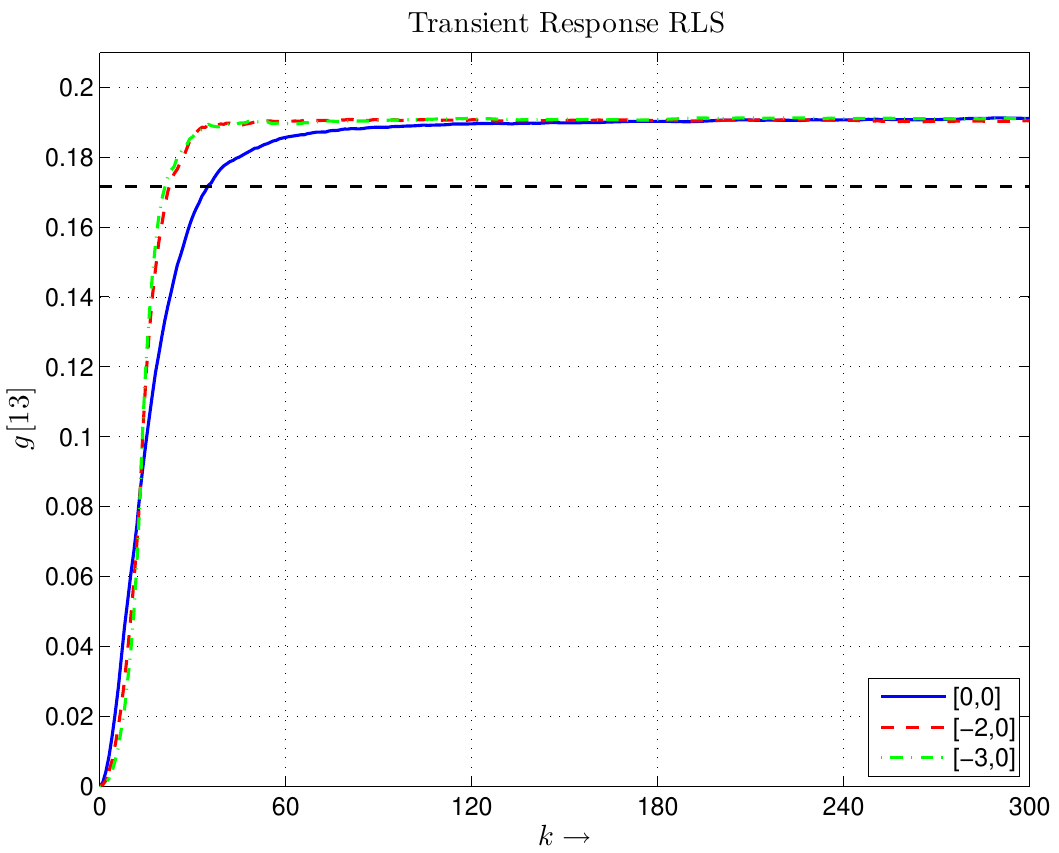}
	\caption{Initial transient response for channel~$3$ with initialization
		$(a,b) = \{(0,0),\,(-2,0),\,(-3,0)\}$.}\label{\figdir/Channel3.pdf}
\end{figure}
\vspace{1em}

\paragraph{Channel $H_3$}

The channel~3 is an approximation of an allpass channel. For an allpass
the eigenvalues are exactly unity. The infinite impulse response
has been limited to $30$ tabs.
Hence we obtain for the approximation
the extremal eigenvalues 
$\lambda_{max} = 1.0444, \,  \lambda_{min} = 0.9348$
which are close to the expected values. We expect here the smallest
gain in settling time from a suitable distribution
of the initial eigenvalues, which is confirmed  by \fig{\figdir/Channel3.pdf}.

\begin{figure}
	\centering
	\includegraphics[width=0.99\columnwidth,keepaspectratio]{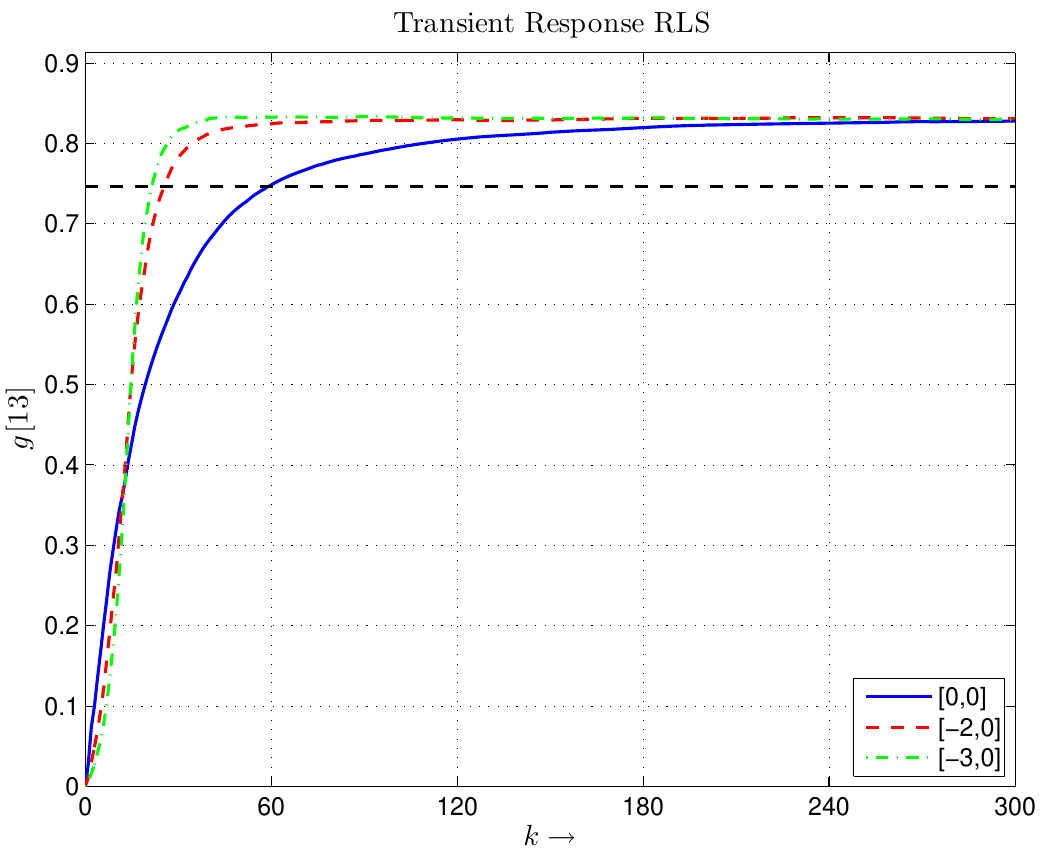}
	\caption{Initial transient response for channel~$4$ with initialization
		$(a,b) = \{(0,0),\,(-2,0),\,(-3,0)\}$.}\label{\figdir/Channel4.pdf}
\end{figure}
\vspace{1em}

\paragraph{Channel $H_4$}
The channel~4 is a mixed phase channel with poles and zeros depicted in
\fig{\figdir/ChannelMixedPhase.pdf}.
The eigenvalue distribution reads
$\lambda_{max} = 4.4670, \,  \lambda_{min} = 0.0447,\,
\lambda_{max}/\lambda_{min} = 99.84$.
The initial transient responses are depicted in 
\fig{\figdir/Channel4.pdf}. Again a significant improvement 
of the settling time could be achieved, compared with an initialization
by the identity matrix.

\rengSubsection{Conclusion and future work}

This paper analyzes the impact of the Cauchy Interlace Theorem on
a suitable initialization of the RLS algorithm
w.r.t.\ a short settling time.
It can be shown
that a proper initialization of the autocorrelation matrix
without clustered eigenvalues can speed
up the convergence by a factor larger than $2$.
The investigations suggest, that w.r.t.\ fast convergence, the initial
estimates of the eigenvalues of the autocorrelation matrix
shall be spread logarithmically 
within the range $10^{-3} \le \lambda \le 1$.
This result is in contract to existing literature, where it is suggested
to initialize the autocorrelation matrix by equal eigenvalues.
It is shown, in contrary, that clustered eigenvalues are the worst
of all choices.
In future work the impact of Cauchy's Theorem on 
other adaptive filters such as the Kalman
filter will be investigated.

\rengSubsection{Acknowledgements}

This project AMOR~ATCZ203 has been co-financed by the European Union using financial means of the European Regional Development Fund (INTERREG) for sustainable cross boarder cooperation. 
Further information on INTERREG Austria-Czech Republic is available at https://www.at-cz.eu/at.
\mbox{\includegraphics[width=0.06\textwidth]{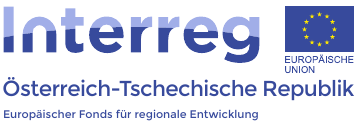} }

\rengSubsection{Data availability}

The datasets generated during and/or analyzed during the current study 
and MATLAB source files
are available from the corresponding author on request.

\printbibliography 

\vspace{0cm}
\begin{center}
	\noindent{\Large\bfseries About the Author}
	\vspace{0cm}
\end{center}

\noindent\textbf{Hans Georg Brachtendorf } graduated in Electrical Engineering from RWTH Aachen, 
Germany in 1989 and obtained the Ph.D.\ degree from the University of Bremen at the
Institute for Electromagnetic Theory and Microelectronics in 1994, 
also in Electrical Engineering, respectively. From 1994-2001 he was an Assistant Professor (C1) also at the University of Bremen 
and obtained the Venia Legendi (Habilitation) 
from the same university in 2001. In 1997-1998 he was 
affiliated with the Wireless Laboratory of Bell Laboratories/Lucent Technologies in 
Murray Hill/New Jersey. 
In 2001 he joined the Fraunhofer Institute for Integrated Circuits in Erlangen, Germany. 
Since 2005 he is full professor at the University of Applied Science
of Upper Austria for System Design and Simulation, Communications and Signal Processing. 


\rengAppendix{Novel proof of the Cauchy Interlace Theorem}\label{sec:apx1}\\
\vspace{1em} \vspace{1em}

\begin{figure}
	\centering
	\includegraphics[width=0.99\columnwidth,keepaspectratio]{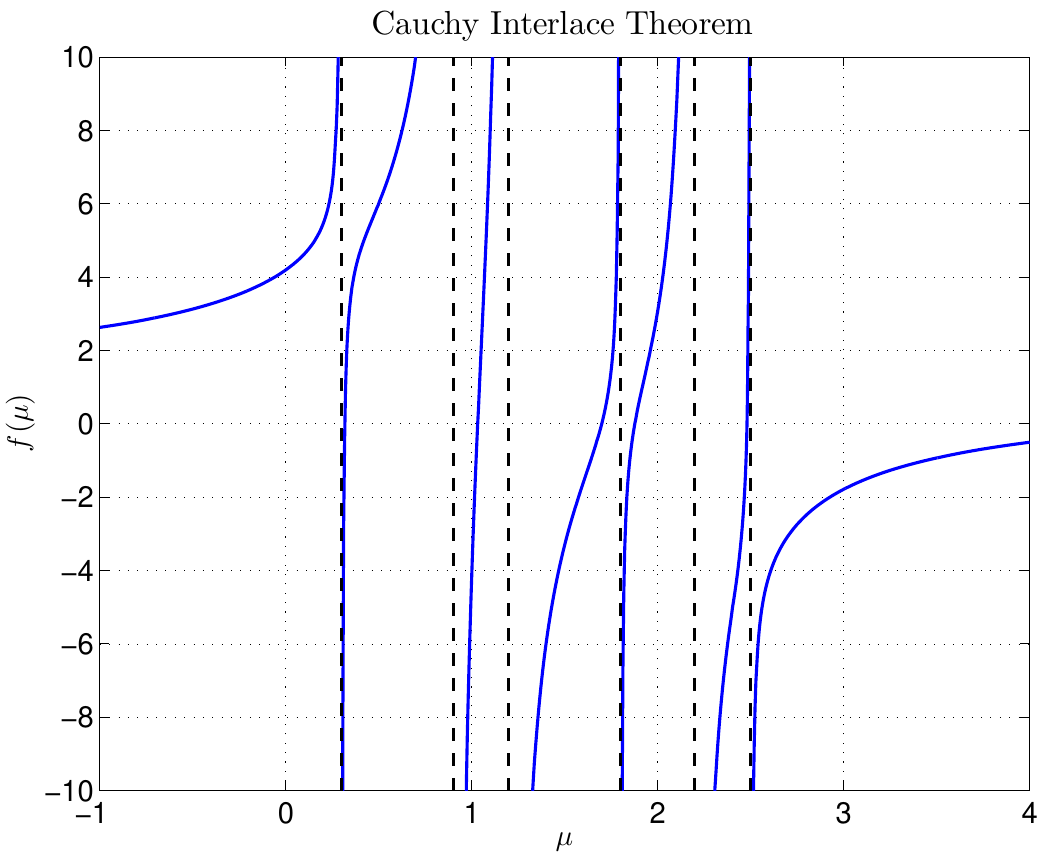}
	\caption{Example 
		for the function $f(\mu)$ with marked eigenvalues $\lambda$.}\label{\figdir/CauchyInterlaceFunction.pdf}
\end{figure}

\begin{theorem} (Cauchy Interlace Theorem)
	Let $A = A^H$ be a Hermitian matrix of dimension $N \times N$ 
	with real eigenvalues $\lambda_i$ sorted in ascending order.
	Moreover, let $u$ be a complex column vector of length $N$. 
	Consider the rank one updated matrix $\tilde{A} = A + u\, u^H$ with eigenvalues
	$\mu_i$ in ascending order. The eigenvalues of $A$ and $\hat{A}$ interlace, i.e.\
	\[
	\lambda_i \le \mu_i \le \lambda_{i+1}
	\]
\end{theorem}

\begin{theorem*}
	Let $Z$ be the matrix of eigenvectors of $A$ such that
	\[
	\Lambda = Z^H\,A\,Z
	\]
	with the eigenvalues in ascending order on the diagonal of
	$\Lambda$. Furthermore, let $v = Z^H\,u$.
	By similarity transform one obtains $\hat{A} = Z^H\,\tilde{A}\,Z = \Lambda + v\, v^H$. 
	$\tilde{A}$ and
	$\hat{A}$ have the same eigenvalues
	\[
	M = \diag\{\mu_1,\dots, \mu_N\}
	\]
	also ascending.
	Assume, that the $i$~th component of $v$ is identical to zero. 
	Then obviously $\mu_i = \lambda_i$. Let this be
	the case for $m \le N$ eigenvalues.  
	The interlacing property of the remaining $N-m$ eigenvalues can be 
	shown as follows\footnote{For ease of presentation we keep for the submatrices
		the notations $\tilde{A},\,\hat{A},\,\Lambda$ etc.}. 
	Assume first that
	all eigenvalues of $\Lambda$ are distinct. Consider the matrix pencil
	\[
	\Lambda - \mu\,I + v\, v^H = D +  v\, v^H
	\]
	with diagonal matrix
	$D = \Lambda - \mu\,I$. For $\mu \in \R,\,  \mu \neq \mu_i,\,i = 1,\dots,N-m$ 
	the inverse matrix exists and can be evaluated
	by the Sherman-Morrison-Woodbury formula, i.e.\
	\[
	(D + v\, v^H)^{-1} = D^{-1} - D^{-1}\,v 
	\cdot c \cdot v^H\,D^{-1},\quad c = \frac{1}{1 + v^H\,D^{-1}\,v}
	\]
	if $D$ is regular and $c$ bounded. Due to the assumption  
	$\mu_i \neq \lambda_i$ the diagonal matrix $D(\mu)$ is regular in 
	an environment of
	$\mu_i$. Consider therefore the zeros of
	\[
	f(\mu) := 1 + v^H\,D^{-1}\,v
	\]
	Expanding $f(\mu)$ leads to
	\[
	f(\mu) = 1 + \sum_{i=1}^N v_i^2\, \frac{1}{\lambda_i - \mu}
	\]
	with the properties
	\begin{align*}
		\lim_{\mu \rightarrow \pm\infty} f(\mu) & = 1\\
		\lim_{\mu \rightarrow \lambda_i+} f(\mu) &= -\infty\\
		\lim_{\mu \rightarrow -\lambda_i-} f(\mu) &= \infty\\
	\end{align*}
	Therefore $f(\mu)$ has $N-m$ zero crossings, the eigenvalues of $\tilde{A}$. 
	Hence the eigenvalues interlace $\lambda_i \le \mu_i$.
	An example is shown
	in \fig{\figdir/CauchyInterlaceFunction.pdf} with marked eigenvalues (singularities of $f$).
	The zeros are the eigenvalues $\mu_i$.
	
	Now the case of two fold identical eigenvalues $\lambda_{j-1} = \lambda_{j}$ is
	considered next. The generalization to $l$ fold identical eigenvalues is straightforward.
	Let $e_j$ be the $j$~th canonical vector and $\Delta\lambda < \lambda_{j+1} - \lambda_j$.
	Consider the perturbed diagonal 
	matrix $\hat{\Lambda} = \Lambda + \Delta\lambda\cdot e_j$. From the proof above it follows
	that the eigenvalues of $\hat{\Lambda} + v\,v^H$ interlace, i.~e.\
	$\lambda_{j-1} \le \mu_{j-1} \le \lambda_{j-1} + \Delta\lambda \le \mu_j 
	\le \lambda_{j+1}$.
	For ${\Delta\lambda\rightarrow 0}$ one concludes
	that $\lambda_{j-1} = \mu_{j-1}$. 
\end{theorem*}

\end{multicols}

\end{document}